# Magmatic sulfides in the porphyritic chondrules of EH enstatite chondrites.

Laurette Piani[1,2*], Yves Marrocchi[2], Guy Libourel[3] and Laurent Tissandier[2]

[1] Department of Natural History Sciences, Faculty of Science, Hokkaido University, Sapporo, 060-0810, Japan

[2] CRPG, UMR 7358, CNRS - Université de Lorraine, 54500 Vandoeuvre-lès-Nancy, France

[3] Laboratoire Lagrange, UMR7293, Université de la Côte d'Azur, CNRS, Observatoire de la Côte d'Azur,F-06304 Nice Cedex 4, France

*Corresponding author: Laurette Piani (laurette@ep.sci.hokudai.ac.jp)

**Abstract**

The nature and distribution of sulfides within 17 porphyritic chondrules of the Sahara 97096 EH3 enstatite chondrite have been studied by backscattered electron microscopy and electron microprobe in order to investigate the role of gas-melt interactions in the chondrule sulfide formation.

Troilite (FeS) is systematically present and is the most abundant sulfide within the EH3 chondrite chondrules. It is found either poikilitically enclosed in low-Ca pyroxenes or scattered within the glassy mesostasis. Oldhamite (CaS) and niningerite [(Mg,Fe,Mn)S] are present in $\approx$ 60 % of the chondrules studied. While oldhamite is preferentially present in the mesostasis, niningerite associated with silica is generally observed in contact with troilite and low-Ca pyroxene. The Sahara 97096 chondrule mesostases contain high abundances of alkali and volatile elements (average $Na_2O$ = 8.7 wt.%, $K_2O$ = 0.8 wt.%, Cl = 7000 ppm and S = 3700 ppm) as well as silica (average $SiO_2$ = 63.1 wt.%).

Our data suggest that most of the sulfides found in EH3 chondrite chondrules are magmatic minerals that formed after the dissolution of S from a volatile-rich gaseous environment into the molten chondrules. Troilite formation occurred *via* sulfur solubility within Fe-poor chondrule melts followed by sulfide saturation, which causes an immiscible iron sulfide liquid to separate from the silicate melt. The FeS saturation started at the same time as or prior to the crystallization of low-Ca pyroxene during the high temperature chondrule forming event(s). Protracted gas-melt interactions under high partial pressures of S and SiO led to the formation of niningerite-silica associations *via* destabilization of the previously formed FeS and low-Ca pyroxene. We also propose that formation of the oldhamite occurred *via* the sulfide saturation of Fe-poor chondrule melts at moderate S concentration due to the high degree of polymerization and the high Na-content of the chondrule melts, which allowed the activity of CaO in the melt to be enhanced. Gas-melt interactions thus appear to be a key process that may control the mineralogy of chondrules in the different classes of chondrite.

**Keywords:** chondrules, sulfides, mesostasis, enstatite chondrite, troilite, oldhamite, niningerite, gas-melt interactions

## 1. Introduction

Among the different types of primitive meteorites, unequilibrated enstatite chondrites (hereafter UEC) are believed to have sampled highly-reducing nebular conditions, as attested to by their high abundances of metallic iron relative to oxidized iron (Sears, 2004). These chondrites are of extreme importance as the isotopic compositions of several major and minor elements are very similar to those found on Earth, and thus the chondrites could have represented a significant part of the Earth's building blocks (e.g. Javoy et al., 2010). In addition, orbital observations of Mercury have revealed that its surface shares a number of chemical and petrographic characteristics with UECs (e.g., Fe-poor and Mg-rich silicates, S-rich minerals) (Nittler et al., 2011; Weider et al., 2012; Evans et al., 2012). UECs are therefore considered to be analogs of Mercury's precursor materials (Zolotov et al., 2013).

UECs are characterized by a peculiar mineralogy, containing unusual phases such as Si-rich Fe-Ni beads, oldhamite (CaS), Mg- and Mn-bearing sulfides (niningerite [(Mg,Fe,Mn)S] and alabandite [(Mn,Fe,Mg)S] in EH and EL enstatite chondrites, respectively) (Keil, 1968).

These phases mainly occur within complex metal-sulfide nodules that, along with pyroxene-rich porphyritic chondrules, represent one of the two main components of UECs (Weisberg and Kimura, 2012). However, Si-rich Fe-Ni metal beads and/or Fe-, Ca-, Mg- and Mn-sulfides are also present in the chondrules (Rubin, 2010; Weisberg and Kimura, 2012), in association with silica, low-Ca pyroxenes (Lehner et al., 2013) and/or linked to relatively large glassy mesostasis pockets (Ikeda, 1989; Piani et al., 2013). The occurrence of these mineral phases in the chondrules is suggestive of specific formation conditions characterized by enhanced sulfur fugacity (Lehner et al., 2013). Sulfides in silica-rich chondrules of EH chondrites have therefore been interpreted to be the result of interactions between an external S-rich gas and ferromagnesian silicates (Lehner et al., 2013). Based on physicochemical analyses, Lehner et al. proposed that the sulfidation process took place at 1400-1600 K in a H-depleted gaseous reservoir enriched in S and C. Under these conditions, the chondrules should have been partially molten and EH chondrite chondrule melts should therefore have recorded high partial pressures of sulfur *via* gas-melt interactions. This process has recently been reported in CV chondrites, where iron sulfide formation was demonstrated to have occurred through complex high-temperature processes of sulfur solubility within the chondrule melt followed by sulfide saturation (Marrocchi and Libourel, 2013). For UEC chondrules, it is conceivable that both processes played a significant role in establishing the diversity of sulfides observed within chondrules. However, UEC chondrule mesostases have received relatively little attention (with the exception of Grossman et al., 1985, Schneider et al., 2002) despite the fact that they represent potential thermochemical sensors of the surrounding gas in the chondrule-forming regions (Cohen et al., 2000; Ozawa and Nagahara, 2001; Libourel et al., 2006; Alexander et al., 2008; Marrocchi and Libourel, 2013, Marrocchi & Chaussidon, 2015). In this context, we report a systematic petrographic and mineralogical survey of sulfides in the porphyritic chondrules of Sahara 97096 (EH3) in order to better establish the conditions under which sulfides and chondrules formed. Sahara 97096 was chosen because of its highly unequilibrated character (Grossman, 1998; Weisberg and Prinz, 1998; Piani et al., 2012; Lehner et al., 2014) as attested to by the presence of $^{16}$O-rich olivines in isotopic disequilibrium with other chondrule phases (Weisberg et al., 2011). In addition, Raman spectroscopy on organic matter indicates that Sahara 97096 has been only mildly metamorphosed (i.e., 3.1-3.4; Quirico et al., 2011). Furthermore, the presence of layered metal-sulfide nodules, the FeS zoning in the niningerite (Lehner et al., 2010), and the noble gas characteristics (Huss and Lewis, 1994) suggest a peak metamorphic temperature of <500°C and limited sulfide remobilization. These characteristics make Sahara 97096 a good candidate for deciphering the processes and conditions that led to the formation of sulfides within EH chondrite chondrules, and by extension, the settings of chondrule-forming events.

## 2. Material and methods

We examined 65 chondrules, including fragments, in one thick section (#3519) and one thin section (#9715) of Sahara 97096 provided by the Natural History National Museum in Paris. Modal abundances and chemical compositions were determined in 17 of the chondrules (6 characterized by porphyritic olivine pyroxene (POP) textures and 11 by porphyritic pyroxene (PP) textures; size ranging from 240 to 1300 μm with a mean size of 530 ± 260 μm). These 17 chondrules contain large (at least a few microns in size) Al-rich mesostasis-like areas, allowing us to investigate the chemical composition of the mesostasis. Backscattered electron (BSE) imaging and chemical mapping using electron dispersive X-ray spectroscopy were performed on JEOL JSM-6510 and JEOL JSM 7000F scanning electron microscopes at CRPG (Nancy, France) and Hokkaido University (Japan), respectively. X-ray maps of Si, Al, Mg, Ca, S, Fe, Ni distributions were used to identify the major phases (low-Ca pyroxene, olivine, silica, high-Ca pyroxene, kamacite, troilite, oldhamite, niningerite and mesostasis). Based on these maps, each chondrule was reconstructed in false colors using Adobe Photoshop®, with a different color used for each major phase, and the modal abundance of each phase was then determined. When crystal sizes were large enough (greater than a few microns in size), modal abundances were also determined for high-Ca pyroxenes in the mesostasis. If not, the high-Ca pyroxene crystallites were included in the total mesostasis surface percent.

Chemical compositions were determined using the CAMECA SX-five electron probe microanalyzer (EPMA) at Université Pierre et Marie Curie (Paris) with a 15 keV electron beam and a wavelength-dispersive X-ray spectrometer. For the determination of mesostasis compositions, analyses were performed on 14 chondrules with a 10 nA electron beam. Care was taken to select areas of mesostasis that contained no micro-crystallites. In addition, EPMA measurements were performed over large scanning areas (from 3 to 30 μm) to avoid the issue of the sub-micron size heterogeneity of some of the mesostasis. Following previous studies (e.g., Wallace and Carmichael, 1992; O'Neill and Mavrogenes, 2002; Liu et al., 2007), we used the VG-2 basaltic glass (Jarosewich et al., 1980) as a reference sample for the mesostasis analyses. For the sulfur content, we measured a value of 1472 ± 58 ppm, which is consistent with values reported in previous studies (e.g. 1403 ± 31 ppm in O'Neill and Mavrogenes, 2002). The detection limit for S in glass under our analytical conditions is estimated to be 300 ppm. Silicate composition and metal and sulfide compositions were measured with focused 15keV/40nA and 15keV/10nA electron beams, respectively. The microprobe was calibrated using the following natural and synthetic standards for silicate analyses: diopside (Si, Ca, Mg), orthoclase (Al, K), albite (Na), pyrite (S), $MnTiO_2$ (Mn, Ti), $Cr_2O_3$ (Cr), $Fe_2O_3$ (Fe), NiO (Ni), apatite (P), and scacchite (Cl). For metal and sulfide analyses we used diopside (Si, Ca, Mg), orthoclase (Al), pyrite (Fe, S), $MnTiO_2$ (Mn, Ti), $Cr_2O_3$ (Cr), $Fe_2O_3$ (Fe), NiO (Ni), and apatite (P). Detection limits are listed in Table 1.

## 3. Results

### 3.1. General characteristics of the chondrules

The 65 chondrules range in size from 200 to 1500 μm. Around 75% of these chondrules exhibited a type I PP texture, but type I POP and radial pyroxene chondrules were also observed (Fig. 1). No type I porphyritic olivine (PO) chondrules were identified, but isolated olivine crystals and olivine radial chondrules were occasionally observed. Low-Ca pyroxene and mesostasis were found to be the major phases within the porphyritic chondrules, representing on average 69.0 (± 12.3) vol.% and 16.2 (± 7.3) vol.% (1σ), respectively (Table 1 and S1). The low-Ca pyroxenes are in the form of large euhedral crystals containing poikilitically enclosed olivines and troilites (Fig. 1, 2 and 3), and have compositions close to the enstatite end-member (i.e., FeO ≤ 1.5 wt. %; Table 1 and S1). The glassy mesostasis often contains crystallites of Fe-, Ca- and Mg-sulfides and Ca-rich pyroxenes and was mainly observed between large pyroxene crystals (Fig. 1 and 2). Olivine was observed in all but three of the 17 chondrules (CH17, CH21 and CH61, Fig. 1, Table S1) either as large, cracked, occasionally dusty (i.e., Fe-Ni metal-bearing) crystals surrounded by low-Ca pyroxene and mesostasis (CH25 and CH45, Fig. 1 and 2C) or as small rounded to subhedral grains poikilitically enclosed in low-Ca pyroxene (CH20 and CH39, Fig 1, 2A and 2D). The olivine compositions are close to the forsterite endmember, with FeO contents ranging from 0.57 to 1.92 wt.% (Table 1 and S1).

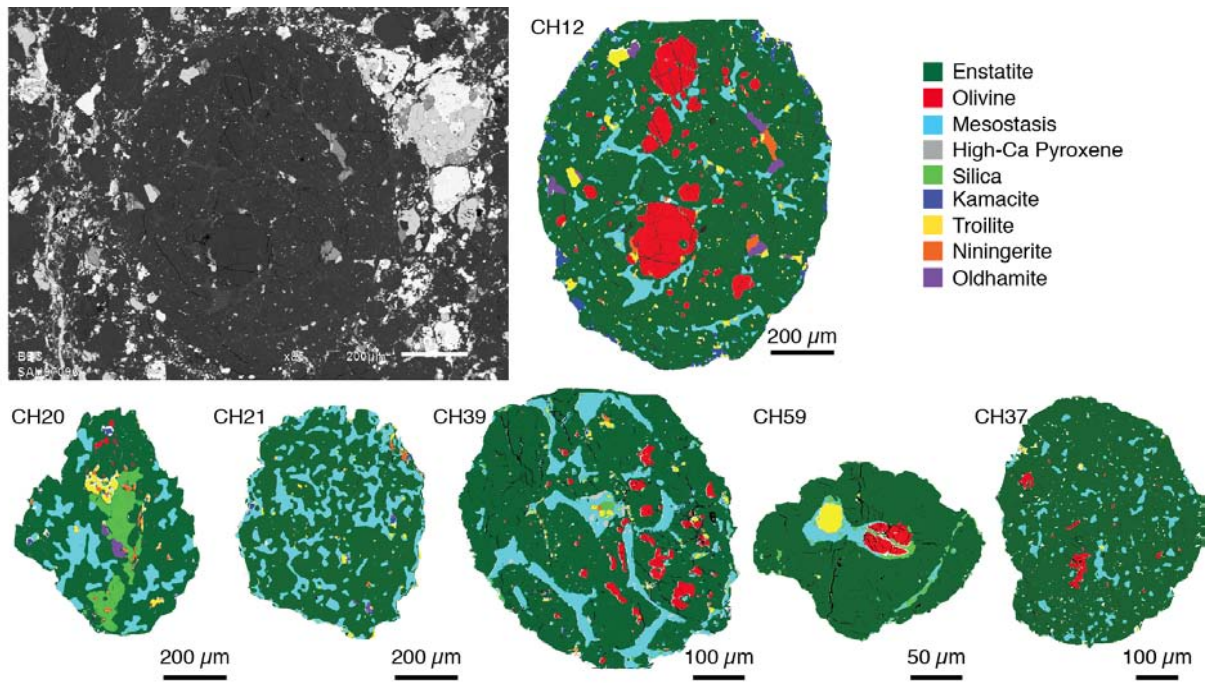

**Fig. 1**. Backscattered-electron (BSE) image of chondrule CH12, and false color maps of CH12 and five other representative chondrules of Sahara 97096 showing their major phases (low-Ca pyroxene, mesostasis, olivine, high-Ca pyroxene, silica, troilite, oldhamite, niningerite and kamacite).

| Average | Mes | En | Ol | CPX | Sil | P.Sil. | | Tr | Nin | Old | Kam |
|---|---|---|---|---|---|---|---|---|---|---|---|
| Vol.% | 16.2 | 69.0 | 8.4 | 0.3 | 2.2 | | Vol.% | 2.1 | 0.4 | 0.5 | 0.7 |
| N | 141 | 113 | 62 | 10 | 4 | 1 | N | 25 | 24 | 3 | 9 |
| $Na_2O$ | 8.68 | b.d.l. | b.d.l. | 0.20 | 0.20 | 0.06 | Na | n.m. | n.m. | n.m. | n.m. |
| σ | 2.20 | | | 0.11 | 0.17 | | σ | | | | |
| $SiO_2$ | 62.8 | 59.2 | 42.8 | 50.9 | 97.3 | 94.0 | Si | 0.03 | 0.09 | b.d.l. | 2.39 |
| σ | 4.25 | 0.90 | 0.72 | 2.52 | 1.62 | | σ | 0.05 | 0.20 | | 0.19 |
| $Al_2O_3$ | 21.4 | 0.53 | 0.05 | 7.02 | 0.98 | 0.11 | Al | b.d.l. | n.m. | b.d.l. | b.d.l. |
| σ | 2.20 | 0.59 | 0.10 | 2.74 | 1.06 | | σ | | | | |
| MgO | 1.88 | 38.2 | 55.3 | 18.8 | 0.19 | 4.60 | Mg | b.d.l. | 24.2 | 0.37 | 0.07 |
| σ | 1.91 | 1.09 | 1.06 | 2.38 | 0.23 | | σ | | 1.85 | 0.02 | 0.08 |
| $K_2O$ | 0.78 | b.d.l. | b.d.l. | 0.03 | 0.04 | 0.03 | K | n.m. | n.m. | n.m. | n.m. |
| σ | 0.52 | | | 0.03 | 0.05 | | σ | | | | |
| CaO | 2.91 | 0.21 | 0.21 | 20.0 | 0.19 | 0.05 | Ca | 0.02 | 0.45 | 53.5 | b.d.l. |
| σ | 2.83 | 0.20 | 0.09 | 2.07 | 0.29 | | σ | 0.04 | 0.17 | 0.20 | |
| S | 0.37 | 0.06 | 0.07 | n.m. | 0.04 | 0.10 | S | 37.1 | 48.5 | 42.9 | b.d.l. |
| σ | 0.16 | 0.12 | 0.01 | | 0.00 | | σ | 0.52 | 1.40 | 0.13 | |
| $TiO_2$ | 0.13 | 0.08 | b.d.l. | 1.56 | b.d.l. | b.d.l. | Ti | 0.15 | b.d.l. | b.d.l. | b.d.l. |
| σ | 0.17 | 0.06 | | 0.35 | | | σ | 0.06 | | | |
| $Cr_2O_3$ | b.d.l. | 0.33 | 0.30 | 0.82 | b.d.l. | b.d.l. | Cr | 2.10 | 0.27 | b.d.l. | b.d.l. |
| σ | | 0.13 | 0.06 | 0.45 | | | σ | 0.67 | 0.24 | | |
| MnO | b.d.l. | 0.11 | 0.11 | 0.20 | b.d.l. | b.d.l. | Mn | b.d.l. | 9.31 | 0.20 | b.d.l. |
| σ | | 0.09 | 0.09 | 0.07 | | | σ | | 1.87 | 0.01 | |
| FeO | 0.23 | 0.90 | 1.05 | 0.30 | 0.19 | 0.25 | Fe | 59.1 | 16.3 | 0.25 | 92.9 |
| σ | 0.21 | 0.45 | 0.39 | 0.21 | 0.07 | | σ | 1.10 | 3.96 | 0.08 | 1.13 |
| NiO | b.d.l. | b.d.l. | b.d.l. | b.d.l. | b.d.l. | b.d.l. | Ni | b.d.l. | b.d.l. | b.d.l. | 3.10 |
| σ | | | | | | | σ | | | | 0.81 |
| $P_2O_5$ | b.d.l. | b.d.l. | b.d.l. | b.d.l. | b.d.l. | b.d.l. | P | b.d.l. | b.d.l. | b.d.l. | 0.03 |
| σ | | | | | | | σ | | | | 0.03 |
| Cl | 0.71 | b.d.l. | b.d.l. | n.m. | b.d.l. | 0.02 | Cl | n.m. | n.m. | n.m. | n.m. |
| σ | 0.68 | | | | | | σ | | | | |
| Co | n.m. | n.m. | n.m. | n.m. | n.m. | n.m. | Co | b.d.l. | b.d.l. | b.d.l. | 0.36 |
| σ | | | | | | | σ | | | | 0.06 |
| Sum | 99.9 | 99.6 | 99.9 | 99.8 | 99.1 | 99.2 | Sum | 98.6 | 99.1 | 97.2 | 98.9 |

**Table 1:** Mean modal and chemical composition of chondrule phases (wt.%) in Sahara 97096. σ is the standard deviation over N measurements of the chemical composition. The modal composition corresponds to the average modal composition of the 17 mapped chondrules. Abbreviations: low-Ca pyroxene (En), olivine (Ol), mesostasis (Mes), Ca-rich pyroxene (CPX), Sil (silica crystals), P.Sil (porous silica), troilite (Tr), niningerite (Nin).Troilite (tr), Oldhamite (Old), low-Ca pyroxene (En), olivine (Ol), mesostasis (Mes), Kamacite (Kam). n.m. = element not measured, b.d.l. = below detection limits. For mesostasis, detection limits are 0.08 wt.% for $P_2O_5$, 0.13 wt.% for $TiO_2$, 0.14 wt.% for FeO and MnO and 0.15 wt.% for $Cr_2O_3$ and NiO. For silicates, detection limits are 0.02 wt.% for Cl, 0.03 wt.% for $K_2O$, $Al_2O_3$, $Na_2O$, $P_2O_5$ and S, 0.04 wt.% for $TiO_2$, 0.06 wt.% for NiO and 0.08 wt.% for $Cr_2O_3$, MnO and FeO. For opaque phases, detection limits are 0.02 wt.% for Al, Ca and P, 0.03 wt.% for Si and Mg, 0.04 wt.% for S and Ti, 0.05 wt.% for Cr, 0.06 wt.% for Mn, 0.10 wt.% for Fe and 0.17 wt.% for Ni and Co.

High-Ca pyroxene occurs as small crystallites within the mesostasis and as overgrowths at the edges of low-Ca pyroxene (CH39 and CH25, Fig. 2C and 2D). Their small size makes it difficult to estimate their modal abundance and chemical composition, however they show a maximum abundance of 3.1 vol.% (CH17; Table S1) and contain a large amount of incompatible elements (e.g., $Al_2O_3$ (7.02 wt.%), $TiO_2$ (1.56 wt.%) and $Cr_2O_3$ (0.82 wt.%); Table 1).

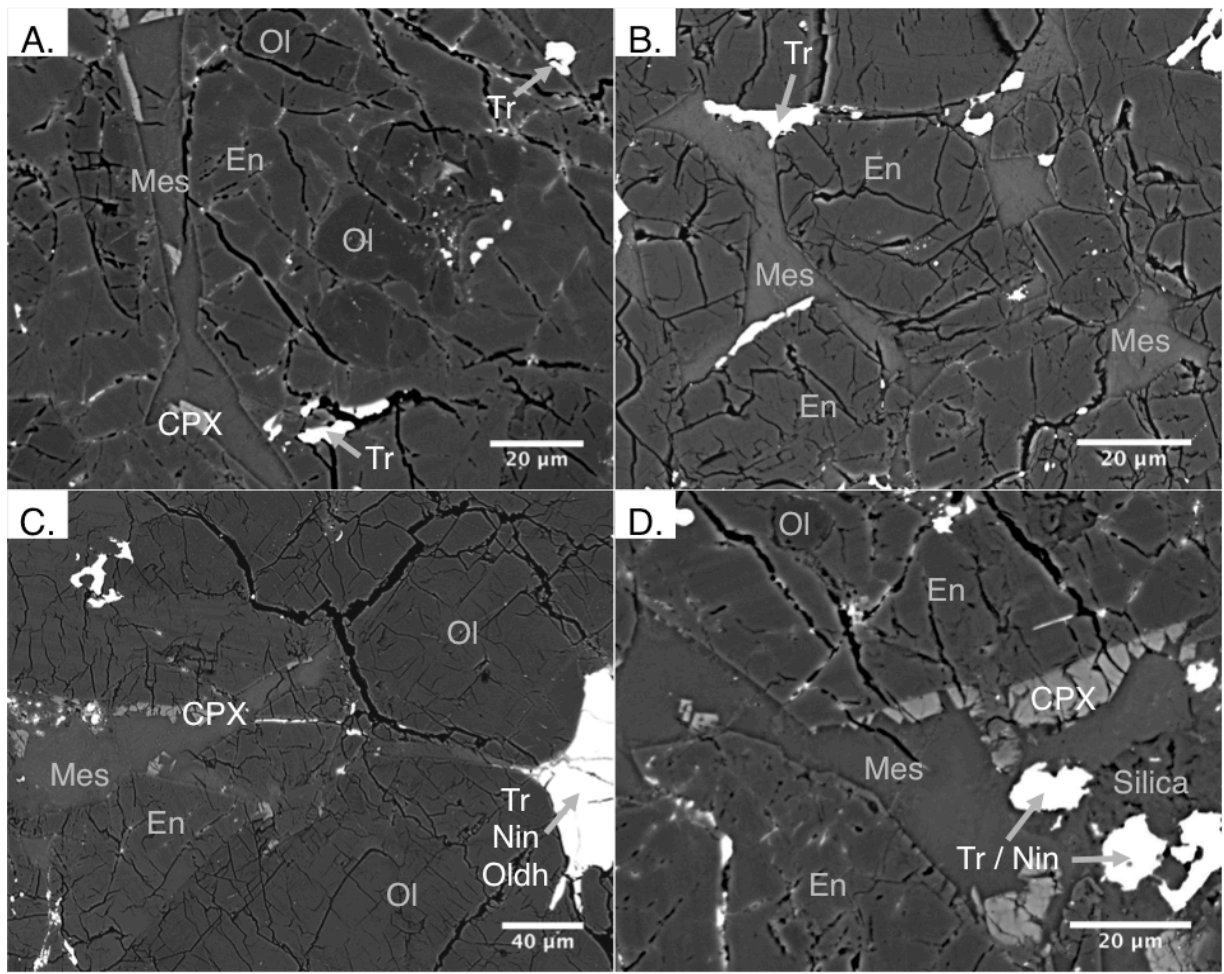

**Fig. 2**. A. Olivine poikilitically enclosed in enstatite pyroxene and surrounded by mesostasis in PP chondrule CH39. Troilite is present in pyroxene and in mesostasis. B. Glassy mesostasis and troilite surrounding enstatite crystals in PP chondrule CH48. C. High-Ca pyroxene and sulfide assemblages in the glassy mesostasis of POP chondrule CH25. D. Olivine enclosed in pyroxene and mesostasis containing high-Ca pyroxene and troilite-niningerite-silica assemblage in PP chondrule CH39. Ol = olivine, En = low-Ca pyroxene (enstatite), CPX = High-Ca pyroxene (clinopyroxene), Mes = mesostasis, Tr = troilite, Old = oldhamite, Nin = niningerite.

Silica occurs in 50% of the chondrules studied in no preferential location and is found either as small crystals in the mesostasis (CH39, CH47 and CH59; Fig. 1 and 2D) or as large porous areas between minerals (CH20, Fig. 1 and 4). The silica-rich porous areas exhibit an irregular shape and vesicular texture (see also Kimura et al., 2005 and Lehner et al., 2013), while the small crystals of silica appear as coherent and homogeneous mineral grains. Of note, the chemical composition of the porous silica may correspond to a mixture of silica and low-Ca pyroxene as indicated by its high MgO content (Table 1 and S1). Although we did not determine the relative proportion of chondrules, matrix and opaque assemblages in Sahara 97096, a previous study reported modal abundances of 22 vol.%, 37 vol.%, and 41 vol.% for matrix, metal-sulfide nodules, and chondrules and silicate clasts, respectively (Lehner et al., 2014).

**3.2. Sulfides in chondrules**

All of the chondrules were found to contain sulfides, mainly in the form of troilite but also as niningerite and oldhamite (Fig. 1, 2, 3 and 4) with average relative abundances of 69, 14 and 17 vol.%, respectively (Table 1 and S1). Sulfide modal abundances vary from 0.5 to 5.4 vol.% for troilite and from 0 to 1.4 and 2.1 vol.% for niningerite and oldhamite, respectively. No caswellsilverite or Cr-bearing sulfides (Grossman et al., 1985) were observed in the chondrules we surveyed. Kamacite can be found as a single phase or in association with sulfides (CH5; Fig. 1 and 3D) within low-Ca pyroxene crystals or on the chondrule edges (Fig.1; Table 1 and S1). Of the 65 chondrules examined in the two sections of Sahara 97096, 22 chondrules contain only troilite, 11 chondrules show $SiO_2$-troilite-niningerite assemblages, 20 chondrules show oldhamite isolated in glassy mesostasis, 13 chondrules present troilite-niningerite-oldhamite associations and 33 chondrules contain troilite-oldhamite, troilite-niningerite or troilite-oldhamite-niningerite assemblages. Twenty-one of the chondrules contain more than one type of these different mineralogical assemblages reported.

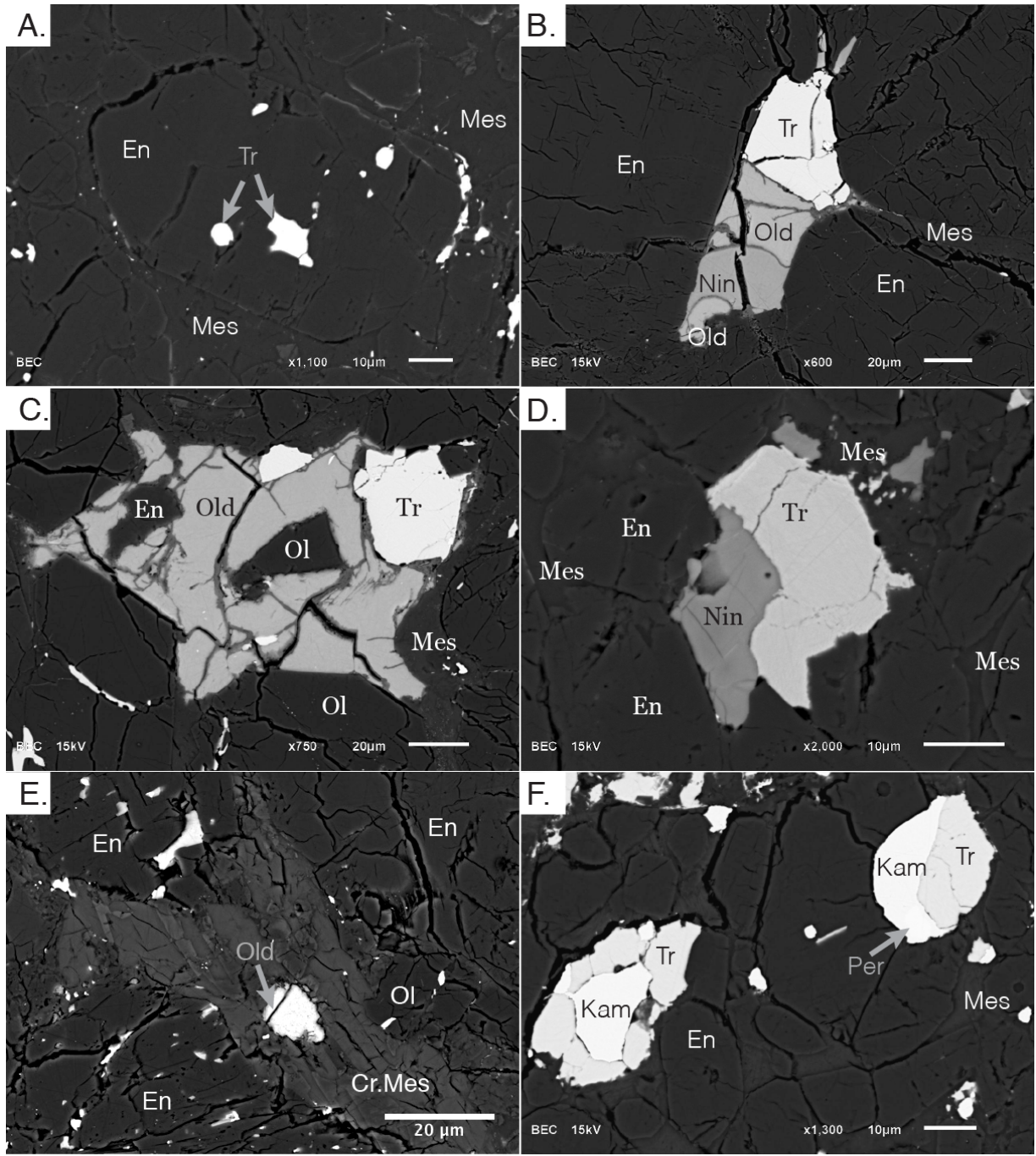

**Fig. 3**: A. Troilite poikilitically enclosed in low-Ca pyroxene and surrounded by mesostasis in PP chondrule CH21. B. Troilite-oldhamite-niningerite assemblage in POP chondrule CH25. C. Troilite-oldhamite assemblage with low-Ca pyroxene and olivine enclosed in oldhamite in POP chondrule CH25. D. Troilite-niningerite assemblage in POP chondrule CH43. The sulfides assemblages in B, C and D show infiltrations between pyroxenes and are all connected to the mesostasis. E. Oldhamite in crystallized mesostasis (anorthic plagioclases and high-Ca pyroxene) in POP chondrule CH45. F. Troilite-kamacite-perryite assemblages in the mesostasis of PP chondrule CH5. Tr = troilite, Old = oldhamite, Nin = niningerite, En = low-Ca pyroxene (enstatite), Ol = olivine, Mes = mesostasis, Cr.Mes = crystallized mesostasis, Kam = kamacite, Per = perryite.

Troilites are systematically present within chondrules and are the most abundant sulfide. They occur as small grains scattered in mesostasis or poikilitically enclosed in low-Ca pyroxenes (1-10 μm) (Fig. 1 and 3). Troilites are also observed in complex assemblages surrounded by mesostasis in association with either niningerite and oldhamite (Fig. 3B, C and D) or Fe-Ni metal (kamacite), phosphide (schreibersite) and silicide (perryite) (Fig. 3F). All types of troilite contain Cr and Ti as trace elements (in concentrations of up to 2.1 wt.% and 0.2 wt.%, respectively; Table S1) and some also exhibit daubreelite ($FeCr_2S_4$) exsolution. The chemical compositions of the porphyritic chondrule troilites are similar to those described in (i) silica-bearing chondrules (Lehner et al., 2013), (ii) the fine-grained matrix (Quirico et al., 2011), and in (iii) opaque assemblages of type 3 enstatite chondrites (Lin and El Goesy, 2002; Zhang et al., 1995). Niningerite and oldhamite are present in $\approx$ 60 % of the chondrules studied and present similar chemical compositions (Table 1 and S1) to those reported in opaque assemblages (Lin and El Goresy, 2002) and silica-bearing chondrules (Lehner et al., 2013). They are often associated with troilite and are surrounded by mesostasis or connected to mesostasis channels between silicates (Fig. 3). These assemblages present liquid or ameboidal shapes with infiltrations along silicate grain boundaries (Fig. 3B, C and D). Silicate inclusions are sometimes present within the oldhamite, as previously reported by Hsu (1998) (Figure 3C). Oldhamite can occasionally be observed as a single phase within the mesostasis (e.g. CH45, Fig. 3E, or CH25 and CH26, Table S1). In contrast, niningerite is frequently found in close association with low-Ca pyroxene, silica and troilite in the silica-rich areas of porphyritic chondrules (CH20 and CH53, Fig. 4; similar assemblages were reported in Lehner et al., 2013). Some of these chondrules (e.g., CH20; Table S1) contain large pools of silica (10.5 vol.% of the chondrule, Table S1), usually related to Al-rich mesostasis areas, in which troilite and niningerite are imbricated (Fig. 1, 4A and 4B). In these assemblages, the niningerite is in direct contact with the low-Ca pyroxene and/or silica but not with the olivine or Al-rich mesostasis (Fig. 4).

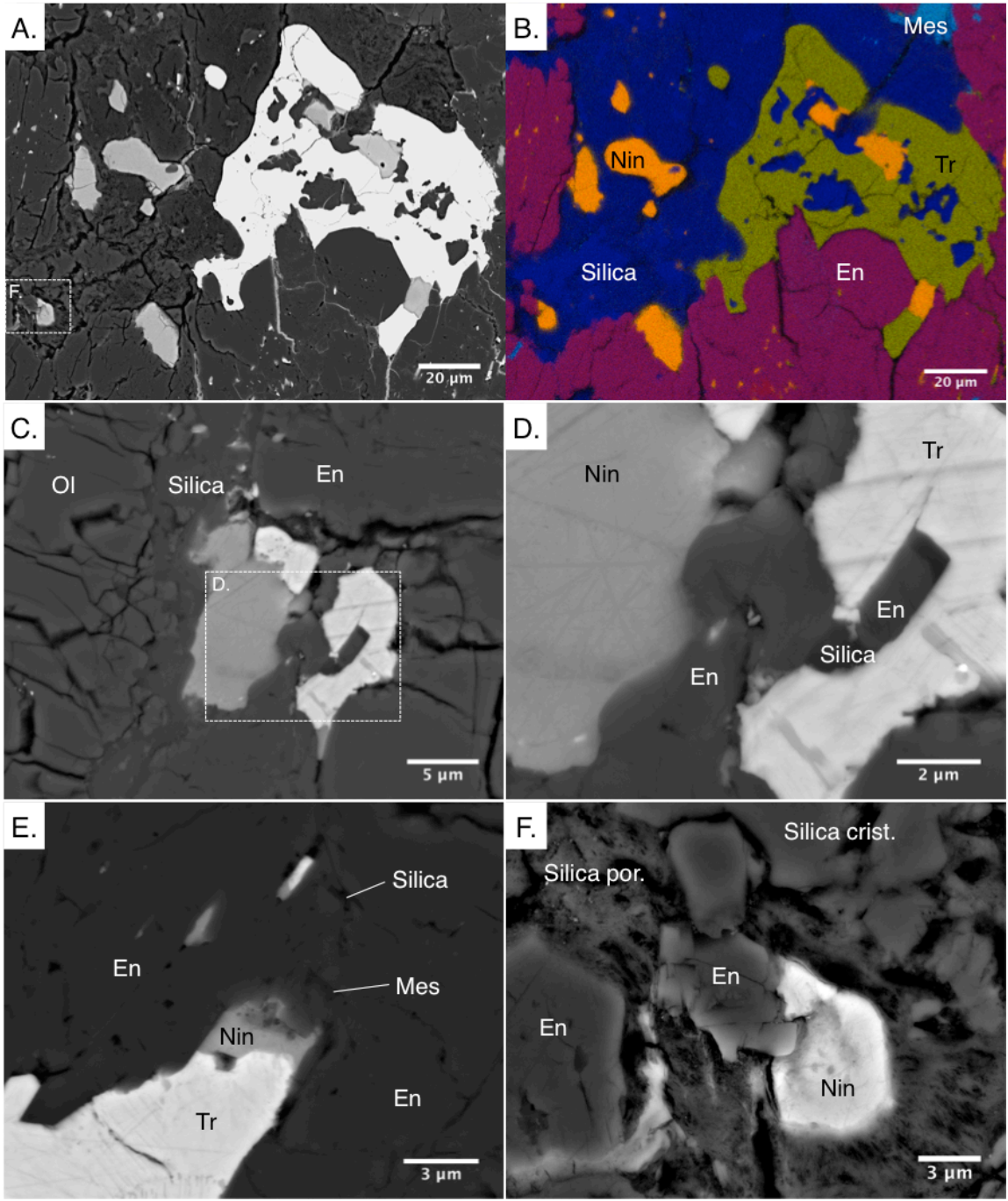

**Fig. 4**: Silicate-silica-sulfide assemblages in Sahara 97096 chondrules. A. BSE image of a large pool of silica (both porous and crystallized) associated with troilite and niningerite in the center of PP chondrule CH20. B. EDS map of A showing Mg in red, Si in dark blue, S in yellow and Al in light blue. C. BSE image in POP chondrule CH53. The rectangle shows the area enlarged in Fig. 4D. The olivine is in contact with silica but not with the sulfides. D. Silica and low-Ca pyroxene are intimately associated with troilite and niningerite assemblages. E. Troilite and niningerite assemblages in low-Ca pyroxene associated with small silica and mesostasis inclusions in chondrule CH20. F. Enlargement of the dotted rectangle shown in A: Low-Ca pyroxene, niningerite and porous and crystalline silica in a large silica pool in CH20. Tr = troilite, Nin = niningerite, En = low-Ca pyroxene (enstatite), Mes = mesostasis, Silica Cr. = crystallized silica, Silica por. = porous silica.

Troilite and troilite-kamacite assemblages, sometimes associated with perryite and schreibersite, are also occasionally found in the mesostasis of chondrules (CH20, Fig. 1 and CH5, Fig. 3F). They appear as globules and, unlike the multi-sulfide assemblages, do not display infiltrations along the silicate grain boundaries. Kamacite in these assemblages contains 2.4 and 3.1 wt. % percent of silicon and nickel, respectively (CH5 and CH20, Table 1 and S1).

### 3.3. Chemical composition of the mesostasis

A large range of average mesostasis compositions was calculated for the Sahara 97096 chondrules (in wt.%): $56.29 < SiO_2 < 69.54$, $18.20 < Al_2O_3 < 23.39$, $5.45 < Na_2O < 11.99$, $0.16 < CaO < 7.88$, $0.48 < MgO < 5.63$, $0.35 < K_2O < 1.87$, $0.14 < FeO < 0.63$ and $TiO_2 < 0.39$ (Table 1 and S1). The mesostasis as a whole is enriched in moderately volatile elements such as silicon, sodium and potassium compared to other chondrite mesostases: Kainsaz (CO), Meteorite Hills 00426 (CR), and Meteorite Hills 00526 (L) (data from Berlin, 2009; Berlin et al., 2011) and Vigarano (CV) (data from Marrocchi and Libourel, 2013) (Fig. 5). Moreover, the Sahara 97096 chondrule mesostases contain significant amounts of volatile elements such as sulfur (mean = 3700 ± 1600 ppm) and chlorine (mean = 7100 ± 6800 ppm) (Table 1). The ranges of average mesostasis compositions for chondrules (in wt.%) are $0.23 < S < 0.50$ and $0.25 < Cl < 1.56$. Isolated measurements show up to 9000 ppm and 3.5 wt.% for S and Cl, respectively (Table S1). Some chondrules also display a chemically heterogeneous mesostasis, as highlighted by the large standard deviations calculated over n measurements in a single chondrule (Table 1 and S1), but no clear chemical zoning was observed from the edges to the cores of chondrules. These data are consistent with previous reports of the mesostasis compositions of EH3 chondrules (Grossman et al., 1985; Schneider et al, 2002).

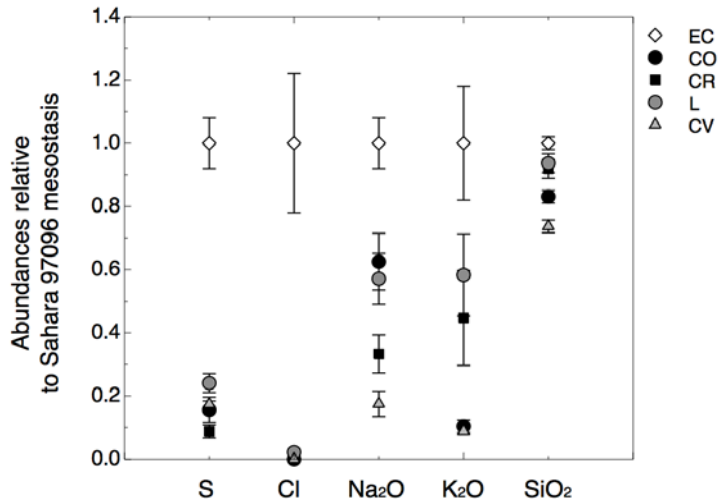

**Fig. 5**: Comparison of the abundance of volatile and moderately volatile elements in the mesostases of Sahara 97096 chondrules (EC) with other chondrites: Kainsaz (CO), Meteorite Hills 00426 (CR), Meteorite Hills 00526 (L) (Berlin, 2009) and Vigarano (CV) (Marrocchi and Libourel, 2013). All other chondrite mesostases have lower volatile and moderately-volatile contents relative to the mean abundance in Sahara 97096 mesostases.

Using the respective proportions of chondrules and matrix in Sahara 97096 (Lehner et al., 2014) and the abundances of the volatile elements in the different chondrule phases (Table 1), we can estimate their distribution at a bulk scale. Sulfur is mainly contained in the matrix and metal-sulfides nodules, while the chondrule mesostases do not contribute significantly to the total S budget. In contrast, the bulk sodium content is almost equally distributed between the chondrule mesostases (60%) and in the matrix (40%). It appears that the chondrule mesostases likely represent the main K-bearing phase even though K-bearing sulfides also contribute significantly to the K total budget ($\approx$ 20%). Although not reported by Lehner et al. (2014), chlorine appears to be distributed among different phases, such as chondrule mesostases, Cl-bearing sulfides (e.g. djerfisherite) in metal-sulfide nodules and porous silica associated with glass-free chondrules (Lehner et al., 2013; this study). It thus appears that, in addition to the matrix and the metal-sulfide nodules, the chondrule mesostasis contributes significantly to the volatile budget of Sahara 97096.

## 4. Discussion
### 4.1. Magmatic origin of sulfides
#### 4.1.1. Troilite

Sahara 97096 chondrules show the systematic presence of sulfides, mostly in the form of troilite (Fig. 1, Table 1 and S1). As this chondrite is considered to be one of the most primitive enstatite chondrites (Weisberg and Prinz, 1998; Quirico et al., 2011; Piani et al., 2012; Weisberg and Kimura, 2012), it is unlikely that the chondrule troilite was formed by redistribution during thermal metamorphism on the parent body (Hewins et al., 1997; Lauretta et al., 1997) or via hydrothermal alteration (Krot et al., 2004). This is confirmed by the systematic occurrence of well-preserved glassy mesostases and Fe-Ni metal beads in porphyritic chondrules (Fig. 1 and 2). Although Sahara 97096 exhibits localized shock features (Rubin and Scott, 1997, Quirico et al., 2011), no molten sulfide veins were observed within the chondrules. Taken together, these observations strongly suggest that chondrule troilite formation was not related to any post-accretion secondary process but was more likely linked to the high temperature chondrule-forming event(s) as previously proposed for sulfides found in other chondrite types and groups (e.g. Kong et al., 2000; Hezel et al., 2010; Marrocchi and Libourel, 2013; Schrader et al., 2016). Thus, troilite could have been present within the EH3 chondrite chondrule precursors and have survived the chondrule-forming high-temperature events. However, experimental and theoretical studies have shown that troilite should have been vaporized under the thermal conditions expected for chondrule formation, even with (i) extremely fast cooling rates (Hewins and Connolly, 1996; Hewins et al., 1997; Hewins et al., 2005) or (ii) very high dust densities with solid/gas enrichments in excess of $10^5$–$10^6$ (Alexander et al., 2008; Fedkin et al., 2012; Alexander and Ebel, 2012; Marrocchi and Libourel, 2013). These conclusions are supported by calculations that show that large troilite inclusions would have reached the surface of chondrules and been evaporated within a short period of time (a few seconds or less; Uesugi et al., 2005).

Among the different sulfides found in Sahara 97096 chondrules, troilite is by far the most abundant, with modal abundances ranging from 0.5 to 5.4 vol.% (Table 1 and S1), and, unlike other sulfides, it is systematically present within chondrules (Table S1). In contrast with other sulfides, troilite is the only sulfide to be enclosed in low-Ca pyroxenes and is systematically found in chondrule glassy mesostases. Because the troilites in the most abundant POP and PP chondrules in Sahara 97096 are poikilitically enclosed in low-Ca pyroxenes (e.g. Fig. 3A), this suggests that the troilite co-crystallized with the low-Ca pyroxenes during high-temperature events. As no other sulfide is found enclosed in low-Ca pyroxene, troilite can be considered to be the first sulfide to have formed in the Sahara 97096 chondrules.

As suggested by Marrocchi and Libourel (2013) for CV chondrules, gas-melt interactions with high partial vapor pressures of sulfur could explain the co-saturation of low-Ca pyroxene and troilite. The frequent association of troilite and low-Ca pyroxene (Fig. 1 and 3) in EH chondrules suggests that the FeS saturation was linked to an increase in the silica activity of the melt ($a_{SiO_2}^{melt}$). This increase in the $a_{SiO_2}^{melt}$ would have resulted from interaction of the melt with the high partial pressure of SiO ($P_{SiO}^{gas}$) of the surrounding environment during the chondrule formation heating event(s). According to the reaction proposed by Marrocchi and Libourel (2013):

$$Mg_2SiO_{4(s)} + FeO_{(melt)} + SiO_{2(melt)} + ½ S_{2(g)} = Mg_2Si_2O_{6(s)} + FeS_{(s)} + ½ O_{2(g)} \qquad (1)$$

an increase in the silica activity of the EH chondrule melt, for a given sulfur fugacity in the surrounding gas, is responsible for (i) olivine dissolution (Fig. 1 and 2A) and then (ii) co-crystallization of low-Ca pyroxene and FeS. The occasional presence of iron-sulfide globules in association with Fe-Ni metal in low-Ca pyroxene (Fig. 3) suggests that the troilite might also have formed from high temperature sulfidation of Fe-Ni metal blebs in the chondrules, as summarized by the following reaction:

$$Mg_2SiO_{4(s)} + Fe_{(s)} + SiO_{2(melt)} + ½ S_{2(g)} = Mg_2Si_2O_{6(s)} + FeS_{(s)} \qquad (2)$$

The assertions above are supported by the calculation of the sulfur content at sulfide saturation, hereafter SCSS, which corresponds to the sulfur content at which an immiscible iron-sulfide liquid (or iron-sulfide solid) separates from the silicate melts (Fincham and Richardson, 1954). Critically, the SCSS calculation predicts that two main factors control sulfur solubility in silicate melts: first, the FeO content of the silicate melt (Fincham and Richardson, 1954; O'Neill and Mavrogenes, 2002), and then, when FeS saturation is reached, the polymerization state of the liquid due to enrichment in network-forming cations such as Si and Al (the SCCS decreasing with increasing polymerization of the liquid; O'Neill and Mavrogenes, 2002). Depending on the model considered, the SCSS might either (i) vary linearly with the iron content of the melt (Li and Ripley, 2009) or (ii) present a U-shape, with an important increase in the S content of the melt in low iron-bearing melts (O'Neill and Mavrogenes, 2002). The sulfur contents of the Sahara 97096 mesostases for the 14 chondrules (from 2300 to 5000 ppm, Table S1) are significantly higher than is predicted in the model of Li and Ripley (<1000 ppm; Li and Ripley, 2009; Fig. 6 and details in Supplement online material). The data are more consistent with the U-shaped model of O'Neill and Mavrogenes, in which the S content of the saturated melt is predicted to be higher than 3000 ppm for FeO < 1 wt. % (Fig. 6). We note however that our data do not perfectly match the U-shaped saturation trend calculated at 1500°C for a melt composition characteristic of UEC porphyritic chondrule (Fig. 6). This discrepancy might linked to a temperature effect as the SCSS is directly dependent on this parameter (O'Neill & Mavrogenes, 2002). In addition, a direct comparison of the sulfur concentration in the EH chondrite mesostases with the SCSS remains difficult due to the lack of experimental data with very low iron contents (< 0.6 wt.%) and the differences in the experimental and chondrule melt compositions (Ariskin et al., 2013; Wykes et al., 2015).

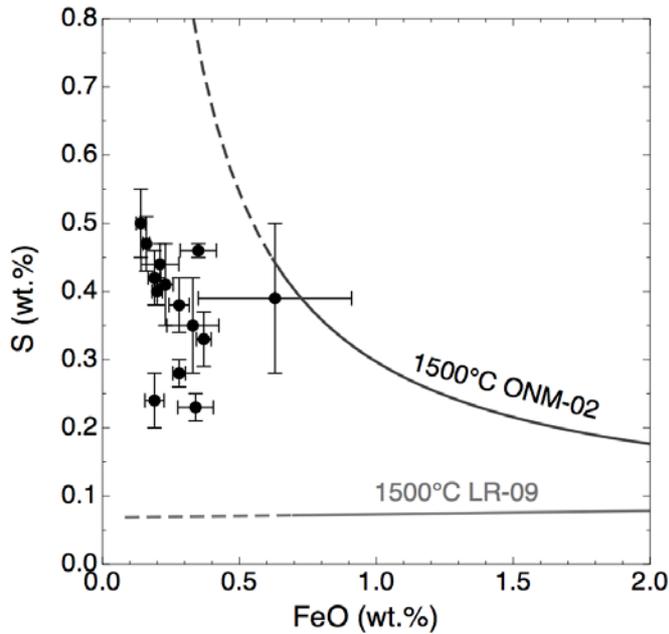

**Fig. 6**: Abundance of sulfur and FeO in the mesostasis of Sahara 97096 chondrules (black dots, Table S1) compared to the sulfur content at sulfide saturation (SCSS) in a silicate melt having the average composition of the chondrule mesostasis of Sahara 97096 (data from Table 1). The SCSS were calculated at 1 bar and 1500°C using the models of Li and Ripley (2009) (black curve LR-09) and O'Neill and Mavrogenes (2002) (ONM-02, grey curve). Details of the calculation are given in Supplementary online materials. Extrapolations of the models for low FeO contents are represented by the dotted lines. Error bars indicate the standard error at 1σ, data and errors for each chondrule are given in Table S1.

Because the troilite is found enclosed in low-Ca pyroxene, the iron-sulfides may have saturated prior to or at the same time as crystallization of the low-Ca pyroxene (the enstatite melting temperature being 1557°C). This therefore requires a saturation temperature for the iron-sulfide melts of about 1500-1600°C. Even though similar magmatic processes are proposed for troilite formation in EH and CV chondrites, it should be noted that the modal abundance of troilite in the EH3 chondrite chondrules (0.5 to 5.4 vol.%; Table 1 and S1) is lower than in Vigarano CV chondrules (up to 15%; Marrocchi and Libourel, 2013). This is likely due to the low iron content of the mesostases of UEC chondrules relative to CV chondrules as iron-poor silicate melts are able to accommodate higher sulfur concentrations before reaching sulfide saturation (O'Neill and Mavrogenes, 2002). The iron content of chondrule melts thus appears to be a key parameter in controlling the mineralogy of chondrules.

### 4.1.2. Niningerite

Unequilibrated EH chondrites contain silica-bearing chondrules with abundant niningerite [(Mg,Fe,Mn)S] associated with troilite, distinguishing them from silica-bearing chondrules in ordinary and carbonaceous chondrites. A striking feature concerning the occurrence of niningerite in Sahara 97096 is the intimate association of niningerite, troilite and low-Ca pyroxene with a silica phase (Fig. 4). Such silicate-silica-sulfide assemblages have also recently been studied by transmission electron microscope observations in silica-bearing chondrules of unequilibrated EH3 chondrites (Lehner et al. 2013). From their petrological and thermodynamical study, Lehner et al. (2013) proposed that niningerite and oldhamite were formed by the sulfidation of ferromagnesian silicates, as previously suggested for silica-rich clasts in the Adhi Kot EH4 chondrite breccia (Rubin, 1983). Their physicochemical analysis of mineral reactions in sulfidized chondrules suggests that molten metal-sulfide assemblages can generate sufficient S vapor to drive sulfidation in a temperature range of 1400-1600 K. The reaction of silicates with the S-rich gas would induce the progressive extraction of Fe, Ca, and Mg into sulfides, with the stoichiometric amounts of silica either reacting with olivine to form enstatite or, when olivine is exhausted, precipitating as free silica (Lehner et al. 2013), according to the following reactions:

$$2\ Mg_2SiO_{4(s)} + S_{2(g)} = 2\ MgS_{(s)} + Mg_2Si_2O_{6(s)} + O_{2(g)} \qquad (3)$$

$$Mg_2Si_2O_{6(s)} + S_{2(g)} = 2\ MgS_{(s)} + 2\ SiO_{2(s)} + O_{2(g)} \qquad (4)$$

$$2\ CaMgSi_2O_{6(ss)} + S_{2(g)} = 2\ CaS_{(s)} + Mg_2Si_2O_{6(ss)} + O_{2(g)} \qquad (5)$$

However, our observations of silica-bearing chondrules show three features that are inconsistent with such reactions: (i) quasi-systematic contact between troilite-niningerite and low-Ca pyroxene while no MgS-olivine contact was observed (Fig. 4, see also Fig. 6-8 from Lehner et al., 2013); (ii) connections between the silica pools and mesostasis (Fig. 4B and 4E); and (iii) higher modal abundances of silica than predicted by reactions (3), (4) and (5) (Fig. 4, CH20 and Lehner et al., 2013). Based on these observations, we therefore suggest an alternative explanation for the origin of silicate-silica-sulfide assemblages. After the co-formation of the low-Ca pyroxene and troilite, further interactions between the SiO- and S-rich gas and the chondrule melts would have destabilized the newly precipitated low-Ca pyroxene and troilite in contact with the melt, resulting in the formation of niningerite and silica assemblages. We therefore propose a reaction in the form of:

$$\tfrac{1}{2}\, Mg_2Si_2O_{6(s)} + FeS_{(s)} + SiO_{2(melt)} = (Mg,Fe)S_{(s)} + 2\, SiO_{2(s)} + \tfrac{1}{2}\, O_{2(g)} \qquad (6)$$

The destabilization could be driven by one or a combination of the following processes: (i) a re-increase in temperature (Skinner and Luce 1971; Andreev et al., 2006); (ii) a decrease in the redox conditions; and/or (iii) an increase in the silica activity of the melt due to the high $P_{SiO_2}^{gas}$ of the surrounding environments. It is worth noting that this process (Eqn. 6) can also explain the high modal abundance of silica found in EH3 chondrules. As suggested by the MgS-FeS phase diagram (Skinner and Luce 1971; Andreev et al., 2006), the position of the MgS-FeS solid solution above the solidus places the lower limit for such a reaction at around 1100°C. Moreover, according to the 30-40 mol. % FeS solubility in niningerite, temperatures as high as 1500-1600°C might be inferred (Andreev et al., 2006). The MnS (Alabandite) contents in niningerite (up to around 20 mol.%, Table S1) and the occurrence of a complete solid solution between MgS and MnS also seem consistent with this range of high temperatures (Shibata, 1926; Andreev et al., 2006).

This melt-assisted reactions can also explain several features observed in EH3 chondrules, for example: (i) the frequent occurrence of niningerite in mesostasis pockets (Fig. 3); (ii) the reverse Fe-Mg zoning in some late crystallized low-Ca pyroxenes (see Fig. 5 of Lehner et al., 2013) due to iron partitioning between the niningerite and mesostasis; and (iii) the preservation of unreacted FeS and low-Ca pyroxene assemblages when the poikilitically-enclosed troilite is shielded from the molten mesostasis by the low-Ca pyroxene (Fig. 3).

### 4.1.3. Oldhamite

Oldhamite can be found in close association with troilite and niningerite in Sahara 97096 chondrules (Fig. 3), but it is also frequently observed as a single phase within the mesostasis (see Fig. 3E and chondrules CH25, CH26 in Table S1). Combined with the noticeable absence of oldhamite in low-Ca pyroxenes, this is consistent with experimental data of Fogel et al. (1996) and McCoy et al. (1999), that show that oldhamite sulfides cannot sustain high temperatures without being melted and dissolved in the silicate melt. This suggests that oldhamite crystallized directly from the melt during the high temperature chondrule-forming event(s), though postdating the formation of the troilite and the low-Ca pyroxene. If so, it may seem surprising that the S content of the co-existing mesostasis is relatively low (< 1 wt% S, see Table 1 and S1) compared to CaS-saturated molten slag, silicate and aluminate melts (Fincham and Richardson, 1954) and to experimental silicate melts produced from the melting of the Indarch EH4 chondrite (Fogel et al. 1996; McCoy et al. 1999), in which S solubility can reach several weight percent.

However, the compositional and structural changes of the melt can explain this apparent inconsistency. Fincham and Richardson (1954) demonstrated that S solubility in melt in reducing conditions can be described in terms of reactions between base metal oxides and $S^{2-}$ species according to:

$$S^{2-}_{(melt)} + \tfrac{1}{2} O_{2(g)} = O^{2-}_{(melt)} + \tfrac{1}{2} S_{2(g)} \qquad (7)$$

They also showed that sulfide capacity can be used to compare the ability of melts of differing composition to accommodate S in their structures:

$$C_s \approx [S] \cdot (P_{O_2} \cdot P_{S_2}^{-1})^{1/2} \qquad (8)$$

where [S] is the sulfide concentration in weight percent and $P_{O_2}$ and $P_{S_2}$ are the partial pressures of $O_2$ and $S_2$, respectively. Essentially, for a given temperature and composition, the higher the sulfide capacity, the greater the ability of a melt to dissolve S. In terms of oldhamite saturation, equation (7) suggests that sulfide capacity depends on the CaO activity in the melt until oldhamite saturation is reached, according to the following reaction:

$$CaO_{(melt)} + \tfrac{1}{2} S_{2(g)} = CaS_{(melt)} + \tfrac{1}{2} O_{2(g)} \qquad (9)$$

Since the S contents of the melt need to rise considerably before oldhamite will precipitate, S solubility in reduced aluminosilicate melts can reach the very high concentrations of several wt. % mentioned above (Fincham and Richardson, 1954; Fogel, 2005). This essentially means that the only oxygen atoms that are readily replaceable by sulfur atoms ($S^{2-}$) are those that are independent of silicon and associated with metal atoms only (i.e., Ca–$O^{2-}$). The sulfide capacity is thus very strongly dependent on the degree of polymerization of the melt. O'Neill and Mavrogenes (2002) proposed that the Fincham-Richardson reaction (7) could be rewritten to take into account such an effect on the structural changes of the melt:

$$S^{2-}_{(melt)} + \tfrac{1}{2} O_{2(g)} + \tfrac{1}{2} SiO_2^{0}{}_{(melt)} = \tfrac{1}{2} S_{2(g)} + \tfrac{1}{2} SiO_4^{4-}{}_{(melt)} \qquad (10)$$

where $SiO_2^0$ and $SiO_4^{4-}$ represent a fully polymerized unit with no non-bridging oxygen and an orthosilicate unit in which Si is coordinated by four non-bridging oxygen atoms, respectively. Accordingly, the more polymerized the melt, the lower its ability to hold sulfide, and the increased polymerization thus facilitates the sulfide saturation. This means that oldhamite saturation can be reached at a lower S content in polymerized melts, such as the Si- and Al-rich melt of EH chondrite chondrules, than in their depolymerized counterparts.

In the case of Sahara 97096, oldhamite saturation in chondrules was very likely helped by additional entry of Na in the molten mesostasis from the gas phase according to a reaction in the form of:

$$2\ Na_{(g)} + \tfrac{1}{2}\ O_{2(g)} = Na_2O_{(melt)} \qquad (11)$$

In producing an almost fully polymerized albitic ($NaAlSi_3O_8$) melt (i.e., after silicate and Fe-Mg sulfide precipitation, see Tables 1 and S1), the entry of sodium would also have been responsible for significant structural changes in the melt. The Na replaces Ca as the charge balancing cations for aluminum in tetrahedral coordination according to:

$$CaAl_2O_4 + Na_2O = CaO + Na_2Al_2O_4 \qquad (12)$$

This is due to the greater stability of the $Na_2Al_2O_4$ component in the melt with respect to $CaAl_2O_4$ from an energetic point of view (see Navrotsky, 1995; Hess, 1995 for review). The addition of $Na_2O$ to an aluminous silicate melt is thus expected to have a similar effect on the $a_{CaO}^{melt}$ as addition of CaO would have on an equi-molar basis (see also Libourel, 1999). Of note, the high activity of Ca in the silicate melt is also supported by the frequent occurrence of high-Ca pyroxenes in mesostasis despite the low concentration of Ca in the melt.

Taken together, the concomitant entries of Na, SiO and S from the gas phase into the chondrule silico-aluminous melt and reactions (9) and (12) favor increases in the degree of polymerization of the silicate melt and the activity of CaO, thereby promoting oldhamite saturation at low S content in the melt according to the following reaction:

$$CaAl_2O_{4(melt)} + Na_2O_{(melt)} + SiO_{(g)} + \tfrac{1}{2}\ S_{2(g)} = CaS_{(s)} + (Na_2Al_2O_4 + SiO_2)_{(albitic\ melt)} \qquad (13)$$

Though examined here in chondrules, most UEC oldhamites are in fact hosted within complex metal-sulfide assemblages located outside the chondrules (Lin and El Goresy, 2002; Gannoun et al., 2011; Weisberg and Kimura, 2012). These are generally considered to be condensation products from the solar nebula on the basis of their high concentrations of Rare Earth Elements (REE) (e.g., 10 to 100 times the mean CI chondrite abundance; Lodders and Fegley, 1993; Gannoun et al., 2011). Similar REE patterns have also been reported within chondrule oldhamites, suggesting that these could also represent nebular condensates that escaped evaporation during the chondrule-forming event(s) (Crozaz and Lundberg, 1995). However, REE enrichments have also been observed for oldhamites within enstatite achondrites (i.e., in aubrites; Dickinson and McCoy, 1997) despite the fact that they should have been erased during the aubrite melting episode (McCoy et al., 1999). Hence, it has been proposed that oldhamite is the crystallization product of an evolved melt enriched in incompatible elements (e.g., REEs, Ca etc.), within both aubrites (Dickinson and McCoy, 1997; Fogel, 2005) and EC chondrites (Hsu, 1998; Jacquet et al., 2015; this study). This assertion is also supported by recent experimental results (Wood and Kiseeva, 2015) that show that in silicate melts with low FeO concentrations, such as the EC chondrule mesostasis (below around 1 wt. %), the partition coefficients of REE between sulfide and melt strongly increase. Hence, the peculiar REE patterns of CaS may not be inconsistent with a magmatic origin for CaS crystallizing from an evolved melt depleted in FeO and enriched in incompatible elements.

It might be argued that the late formation of CaS would induce negative europium (Eu) anomalies in CaS since Eu is preferentially incorporated in FeS. Such negative anomalies are not observed in EH3 enstatite chondrite CaS (Crozaz and Lundberg, 1995; Gannoun et al., 2011). However, the low modal abundance of FeS ($\leq$ 5.4 vol%; Table S1) relative to chondrule glassy mesostases ($\leq$ 16.2 vol%) may not be significant enough to produce Eu anomaly relative to other REE in the chondrule mesostases. Moreover, experimental results have revealed that, unlike the other REE, Eu is preferentially incorporated in CaS rather than in silicate melts (Lodders, 1996). Thus it appears that the presence of the different sulfides in EH chondrite chondrules can be explained by the sequential sulfide saturation from a chondrule melt in interaction with an S-rich gas.

### 4.2. Implications for EH chondrite chondrule formation

The data reported here strengthen the argument for a magmatic origin for sulfides within EH chondrite chondrules in response to reaching sulfur saturation in the chondrule silicate melt and the subsequent sequential crystallization of troilite, niningerite and oldhamite at high temperature during the chondrule-forming event(s). We note, however, that the rare occurrences of other types of sulfides that were reported before within one chondrule (i.e., caswellsilverite and Cr-bearing sulfides; Grossman et al., 1985) are not explained by our model and could correspond to sulfides that were present among the chondrule precursors (Rubin, 2010). However, although we cannot firmly exclude this possibility, this is not supported by the experimental melting of Indarch (EH4) in the range 1000-1400°C, which induced the complete melting of all types of sulfides (Fogel et al., 1996; McCoy et al., 1999). Thus, it appears unlikely that sulfides would have resisted the chondrule-forming event(s), even if the chondrules were only partly melted and cooled quickly (Yu et al., 1996; Rubin, 2010). On the contrary, our petrological observations and chemical analyses support a magmatic origin of sulfides *via* gas-melt interactions under high $PS_{2(g)}$. In such a model, the FeS saturation was reached first and was simultaneous with the crystallization of low-Ca pyroxene. Further addition of S, Si and Na from the gas phase to the melt would have allowed (i) the destabilization of FeS and low-Ca pyroxene, resulting in the formation of niningerite and silica at temperatures potentially as high as 1500-1600°C, and (ii) the saturation of CaS and the crystallization of high-Ca pyroxene in the mesostasis, due to an increase in the calcium activity and in the polymerization degree of the melt. Because of the volatile behavior of sulfur under nebular canonical conditions, the present findings suggest that sulfides in EH chondrite chondrules formed within a non-canonical gaseous environment enriched in sulfur.

In addition to their high S-contents, the Sahara 97096 chondrule mesostases show concentrations of volatile and moderately volatile elements (Na, K, Cl and Si) that are significantly higher than in chondrule mesostasis in any other type of chondrite (Fig. 6, Table 1 and S1; Grossman et al., 1985; Schneider et al., 2002, Alexander et al., 2008; Berlin, 2009; Rubin and Choi, 2009; Marrocchi and Libourel, 2013). The quasi-systematic presence of resorbed or poikilitically-enclosed olivines within low-Ca pyroxenes in the 65 chondrules observed in this study, suggests that olivine can be considered as a precursor material that was out of equilibrium with the mesostasis from the time of the low-Ca pyroxene formation. Low-Ca pyroxenes and troilite assemblages within the mesostases also show destabilization into niningerite and silica (Fig. 4 and Section 4.1.2). These observations are consistent with the REE patterns of silicates in Sahara 97096 chondrules, which show that the mesostases evolved out of equilibrium with the chondrule silicates (Jacquet et al., 2015).

A formation process involving protracted gas-melt interactions for sulfide in EH chondrite chondrules might also be close to that expected for CV type I or CR type II chondrule formation (Marrocchi and Libourel, 2013; Schrader et al., 2015), which would suggest that this unusual process may have been ubiquitous in the chondrule-forming regions. More generally, this result is in line with growing evidence that suggests that chondrule formation took place in an open system under non-canonical conditions with enhanced partial pressures of alkali and volatile elements (Georges et al., 2000; Tissandier et al.; 2002, Krot et al., 2006; Libourel et al., 2006; Ruzicka et al., 2007, 2008; Mathieu, 2009; Hezel et al., 2010; Mathieu et al., 2011; Hewins and Zanda, 2012; Alexander and Ebel, 2012; Fedkin and Grossman, 2013; Marrocchi and Libourel, 2013; Schrader et al., 2013, 2014; Marrocchi et al., 2015, 2016; Marrocchi and Chaussidon, 2015). In addition, Mg-isotopic abundances in bulk chondrule also suggest formation at high total pressure (Galy et al., 2000).

How such conditions were generated, in particular the high partial pressures of alkali and volatile species without generating high oxygen fugacity, remains a key question regarding chondrule formation. Thermodynamic calculations have suggested that the sodium content of type II chondrules could have been established in plumes generated by collision between planetesimals (Fedkin and Grossman, 2013), tough recent studies have questioned this model (Ebert and Bischoff, 2016). However, the dust enrichment expected in impact plumes (i.e., > $10^3$; Fedkin and Grossman, 2013) would have generated oxidizing conditions ($\approx$ IW; Palme and Fegley, 1990; Visscher and Fegley, 2013; Marrocchi et al., 2016) that seem improbable given the peculiar mineralogy of type I EH chondrite chondrules and their inferred extremely reduced conditions (Lehner et al. 2013; Cartier et al., 2014). Dust enrichments are also expected to have occurred during shock waves that produced an evaporation of dust concentrated in the mid-plane of the accretion disk. However, formation during collisions between planetesimals cannot be firmly excluded as variations in dust enrichments might also occur in impact-generated plumes. Nonetheless, the presence of high partial pressures of volatile elements during the formation of EH chondrite chondrules is an important constraint that must be taken into account in the EC chondrule formation models. Chondrules remain our best thermochemical sensors of the surrounding gas for evaluating the prevailing physico-chemical conditions in the chondrule-forming regions of the protoplanetary disk.

## 5. Concluding remarks

The low degree of alteration of the Sahara 97096 EH3 enstatite chondrite has allowed us to characterize the nature and the distribution of sulfides within porphyritic chondrules and the chemical composition of chondrule mesostases. Our data indicate that (i) troilites are ubiquitous within the chondrules and are observed either as inclusions in low-Ca pyroxenes or in glassy mesostases; (ii) oldhamite, niningerite and troilite are frequently observed as assemblages within glassy mesostases; (iii) niningerite can be found in complex assemblages with low-Ca pyroxene, troilite and silica in silica-bearing chondrules; and (iv) oldhamite is sometimes found as a single phase within the mesostasis. In addition, the EH3 chondrite chondrule mesostasis is enriched in alkali and volatile elements compared to chondrule mesostases in other types of chondrites.

Interaction between the molten chondrules and their surrounding gaseous environments is a key factor in explaining the characteristics of the sulfides in EC chondrules and the volatile-rich mesostases (schematic summary in Fig. 7). High partial pressures of S and of SiO would have led to the destabilization of the olivine precursors and the co-saturation of FeS and low-Ca pyroxene. In addition, protracted gas-melt interactions would have induced the formation of niningerite-silica associations *via* the destabilization of the previously-formed FeS and low-Ca pyroxene. The formation of oldhamite occurred *via* the sulfide saturation of Fe-poor chondrule melts at low S concentration due to the high concentration of alkali elements (especially sodium), which allowed the activity of CaO to be enhanced. The late saturation of oldhamite from an evolved silicate melt is an interesting mechanism that might explain its high concentration of incompatible elements such as REE. However, experimental studies are needed to better constrain the partition coefficients of REE between sulfide and silicate melts under the particular conditions of CaS saturation from a FeO-poor silicate melt. In addition this model should be tested by petrographic observations and mineralogical characterizations of other enstatite chondrites.

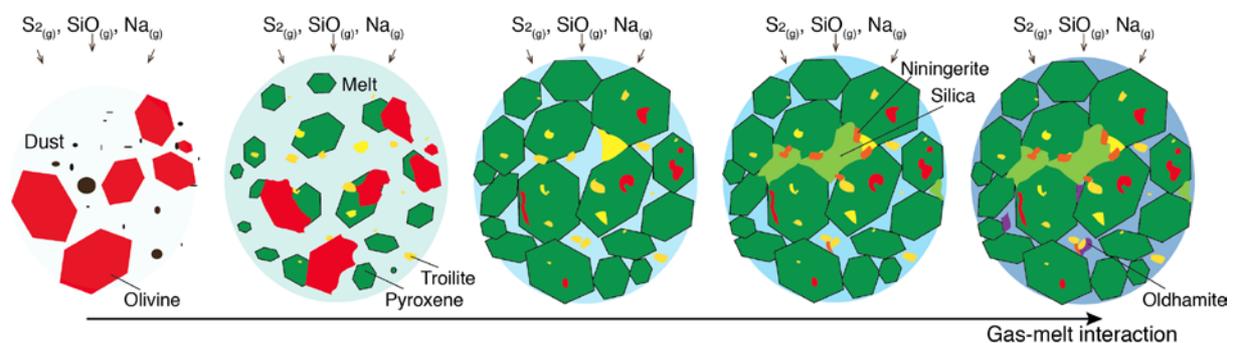

**Fig. 7**: Schematic summary of the formation of sulfides in EC chondrules. Olivine precursors are destabilized in a melt progressively enriched in $SiO_2$ and $S_2$ from the gas, producing the co-saturation of troilite and low-Ca pyroxene. Protracted gas-melt interactions are induced by further incorporation of $SiO_2$ and $S_2$ from the gas to the melt, leading to the formation of niningerite-silica associations *via* the destabilization of the previously formed FeS and low-Ca pyroxene. Oldhamite formed at a later

stage *via* the sulfide saturation of Fe-poor chondrule melts at low S concentration in an evolved silicate melt enriched in sodium and incompatible elements. The darkening of the mesostasis indicates the increase of $SiO_2$ and $Na_2O$ content in the melt due to the gas-melt interaction.

According to our study, the formation of EH chondrite chondrules took place *via* gas-melt interactions characterized by a high partial pressure of volatile elements and SiO. This is in line with previous reports in carbonaceous and ordinary chondrite chondrules of evidence for a high partial pressure of volatile elements in the chondrule-forming regions. This set of results therefore suggests that gas-melt interaction is a common formation mechanism for chondrules, whatever the chondrite class considered. However, in order to explain the differences observed between CV, UOC and UEC chondrules, the specific conditions (i.e., the oxygen fugacity, the partial pressure of volatile elements and the duration of gas-melt interactions) under which the gas-melt interactions took place are not yet fully constrained. Another challenging issue is to better understand how variable high partial pressures of volatile elements and oxygen fugacity might be generated in the chondrule-forming regions. Shock-wave-induced transient evaporation of CI dust or moderately-dust-enriched impact-generated vapor plumes could be among the best candidates.

**Acknowledgements** The authors are grateful to the Museum National d'Histoire Naturelle (Paris, France) for providing the sections of Sahara 97096. François Faure, Delphine Lequin, Romain Mathieu, Michel Fialin, Junji Yamamoto, Kohei Nagata, Maïa Kuga, Noriyuki Kawasaki and Alice Williams are warmly thanked for fruitful discussions, help with data acquisitions and providing other assistance that allowed this work to be completed. We are grateful to Associate Editor Maud Boyet for her helpful review and careful editing. Devin Schrader and two anonymous reviewers are thanked for their thorough and constructive comments. This work has been supported by the Programme National de Planétologie (PI Yves Marrocchi) and by the French National Research Agency, contract ANRBS56-008, project Shocks (PI Guy Libourel). This is CRPG contribution n°2454.